\documentclass
[aps,prb,twocolumn,groupedaddress,draft,showpacs,intlimits,floats]{revtex4}%
\usepackage{eurosym}
\usepackage{bm}
\usepackage[final]{graphicx}
\usepackage{epsfig}
\usepackage{mathptmx}
\usepackage{xspace}
\usepackage{amsmath}
\usepackage{amsfonts}
\usepackage{amssymb}%
\providecommand{\U}[1]{\protect\rule{.1in}{.1in}}
\DeclareMathAlphabet{\mathbfit}{OT1}{cmr}{bx}{it}

\begin{document}
\title{Pinpointing Gap Minima in Ba(Fe$_{0.94}$Co$_{0.06})_{2}$As$_2$ \textit{via} Band Structure
Calculations and Electronic Raman Scattering}
\author{I. I. Mazin$^{1}$}
\author{T. P. Devereaux$^{2}$}
\author{J. G. Analytis$^{2}$}
\author{Jiun-Haw Chu$^{2}$}
\author{I. R. Fisher$^{2}$}
\author{B. Muschler$^{3}$}
\author{R. Hackl$^{3}$}
\affiliation{$^{1}$Code 6393, Naval Research Laboratory, Washington, DC 20375, USA}
\affiliation{$^{2}$Stanford Institute for Materials and Energy Science, SLAC National
Accelerator Laboratory and Dept. of Applied Physics, Stanford University, Stanford, CA 94305, USA}
\affiliation{$^{3}$Walther Meissner Institut, Bayerische Akademie der Wissenschaften, 85748
Garching, Germany}
\date{\today}

\begin{abstract}
A detailed knowledge of the gap structure for the Fe-pnictide superconductors
is still rather rudimentary, with several conflicting reports of either nodes,
deep gap minima, or fully isotropic gaps on the Fermi surface sheets, both in
the $k_{x}-k_{y}$ plane and along the $c-$axis. In this paper we present
considerations for electronic Raman scattering which can help clarify the gap
structure and topology using different light scattering geometries. Using
density functional calculations for the Raman vertices, it is 
shown that the location of the gap minima may occur on loops stretching over a
portion of the $c$-axis in Ba(Fe$_{0.94}$Co$_{0.06})_{2}$As$_2$.

\end{abstract}

\pacs{74.25.nd,74.70.Xa,74.20.Pq,71.15.Mb}
\maketitle

Since the discovery of high temperature superconductivity in the iron
pnictides, identifying the structure of the superconducting order parameter
has remained a tantalizing and elusive problem. Knowledge of the momentum
dependence is a major step towards identifying the interactions that drive the
superconductivity. A variety of experimental techniques probing the low energy
properties of the pnictides in the superconducting state have given
contradictory information on the angular dependence of the superconducting gap
around the Fermi surfaces (FS). Angle-resolved photoemission on NdFeAsO$_{0.9}%
$F$_{0.1}$\cite{Kondo}, K-doped\cite{Ding} and Co-doped\cite{Terashima:2009}
$\mathrm{BaFe_{2}As_{2}}$ have identified rather isotropic superconducting
gaps on both the hole and electron pockets in the Brillouin zone (BZ). On the
other hand, transport and thermodynamic measurements, which do not have
angular resolution, have indicated the presence of nodes in some materials and
small but finite gaps in others\cite{Meingast}. Recently,
angle-resolved thermal conductivity in BaFe$_{2}$(As,P)$_{2}$ have indicated
large gap variations around the Fermi surface, with gap nodes or deep minima
located at $45^{\circ}$ with respect to the principal BZ axes \cite{Matsuda}.

Electronic Raman scattering experiments indicated very small 
gaps located on the
electron Fermi surfaces in Ba(Fe$_{0.94}$Co$_{0.06})_{2}$As$_2$\cite{Muschler:2009}. 
However, due to the relative insensitivity of the
Raman vertices to momentum in the canonical tight binding (TB) scheme the
location of the nodes could not be determined uniquely. The situations is
additionally obscured by the fact that the electronic structure of the 122
materials is inherently more 3D than that of the 1111 pnictides, and
simplified 2D tight binding models can be rather misleading\cite{Kemper}.
Moreover, the crystallographic symmetry of the 122 structure is body-centered
tetragonal (bct), $I4/mmm$, therefore folding down the Fermi surfaces from the
one-Fe Brillouin zone to the two-Fe zone is less trivial than in the 1111
structure. Namely, while in the 1111 structure the electronic pockets at the X
point and Y point in the unfolded zone are related by a simple $\pi/2$
rotation, so that after folding they form e-pockets with the full 4-fold
symmetry, in the 122 structure the symmetry operation connecting the two
unfolded pockets is the same rotation plus a shift along $k_{z}$ by $\pi/c.$
Thus for all $k_{z}$ but $k_{z}=\pi/2c$ the folded FS cuts have only a 2-fold symmetry.

Incidentally, for the same reason the standard method of constructing a TB
Hamiltonian by transforming the Bloch functions into the Wannier functions
leads to a TB model that attempts to describe the outer barrel as a single
band in the unfolded BZ\cite{Kemper}, while in reality it is formed by
anticrossing the X and the Y pockets. In other words, such TB models feature
unphysically and artificially large interlayer hoppings (some interlayer hoppings
in Ref. \onlinecite{Kemper} appear to be more than 50\% of the nearest-neighbor
in-plane hopping), while in reality the scale of the $k_z$ dispersion is set
by the hybridization gaps, and by the difference in size (but not orientation)
between the upper and the lower panel in the Fig. 4 of Ref. \onlinecite{Kemper} .

For this reason, in the 122 systems extracting quantitative information from
the experiment using an infolded 2D tight binding model becomes
questionable, and working in a folded 3D TB model is not much simpler than
straight off first principle calculations. Therefore, in order to gain
more accurate information about the location of gap minima on the Fermi
surface we have performed all-electron band structure calculations, as
described in Ref. \onlinecite{Mazin2008} and numerically computed the relevant Raman
vertices. We have confirmed the result obtained qualitatively by Muschler et
al.\cite{Muschler:2009} that two of the three principal Raman symmetries,
$A_{1g}$ and $B_{2g}$ (all notations in the crystallographic unit cell),
probe, respectively, an average gap and the gap on the electron Fermi surface,
and that for the third symmetry, $B_{1g},$ the matrix elements are mostly
small on all the FSs. However, we found that on the outer electronic FS there
exist \textquotedblleft hot spots\textquotedblright\ where this matrix element
is sizeable, a fact that could not be seen in unfolded TB calculations.
Therefore we can not only distinguish between the hole and the electron FSs in
terms of the gap minima, but can also place them at or near these hot spots
(which appear near the [110] direction).

First, we discuss the so-called \textquotedblleft effective mass
approximation\textquotedblright\ as applied to this compound. The inelastic
light (Raman) scattering cross section in the limit of small momentum
transfers is given in terms of correlation functions of an effective charge
density\cite{RMP}, $\tilde{\rho}=\sum_{n,\mathbf{k},\sigma}\gamma
_{n}(\mathbf{k},\omega_{i},\omega_{s})c_{n,\mathbf{k},\sigma}^{\dagger
}c_{n,\mathbf{k},\sigma},$where $c_{n,\mathbf{k},\sigma}^{\dagger}$
($c_{n,\mathbf{k},\sigma})$ creates (removes) an electron with momentum
$\mathbf{k}$ and spin $\sigma$ from band $n$. In principle, $\gamma$ depends
both on light scattering polarization geometries ${\mathbf{e}^{i(s)}}$ and the
frequencies of incoming (scattered) light $\omega^{i(s)}$. In the limit where
$\omega^{i(s)}$ are much smaller than any relevant interband transition
frequency, the scattering amplitude simplifies into the well-known effective
mass tensor
\begin{equation}
\gamma_{n}(\mathbf{k},\omega^{i,s}\rightarrow0)=\sum_{\alpha,\beta
=x,y,z}e_{\alpha}^{i}\frac{\partial^{2}\epsilon_{n}(\mathbf{k})}{\partial
k_{\alpha}\partial k_{\beta}}e_{\beta}^{s}.
\end{equation}
For tetragonal materials, three different scattering geometries can be
classified according to the point group transformation properties. These are
given as
\begin{align}
A_{1g} &  :~~\gamma_{n}(\mathbf{k})=[\partial^{2}\epsilon_{n}(\mathbf{k}%
)/\partial k_{x}^{2}+\partial^{2}\epsilon_{n}(\mathbf{k})/\partial k_{y}%
^{2}]/2\nonumber\\
B_{1g} &  :~~\gamma_{n}(\mathbf{k})=[\partial^{2}\epsilon_{n}(\mathbf{k}%
)/\partial k_{x}^{2}-\partial^{2}\epsilon_{n}(\mathbf{k})/\partial k_{y}%
^{2}]/2\nonumber\\
B_{2g} &  :~~\gamma_{n}(\mathbf{k})=\partial^{2}\epsilon_{n}(\mathbf{k}%
)/\partial k_{x}\partial k_{y}.\label{basic}%
\end{align}
Thus the curvature of the bands and the light scattering polarization
orientations determine which carriers are involved in light scattering in
different bands and regions of the BZ. While not having the definitive angular
resolution as photoemission to probe single particle properties, Raman
scattering can selectively project effective charge density fluctuations in
different regions of the BZ, uniquely providing $k-$projected dynamical
two-particle properties.

\begin{figure}[t]
\includegraphics[width=0.8\columnwidth]{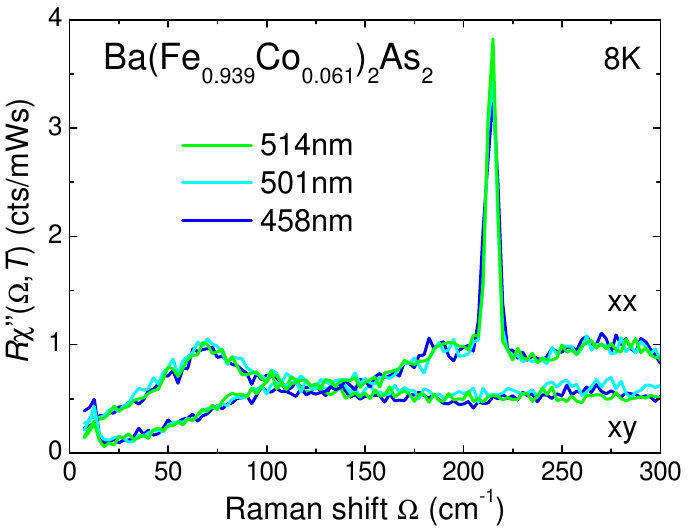}\caption{(Color online) Raman
response for $xx$ and $xy$ polarization for three different excitation lines
as indicated. All spectra have the Bose factor removed. Apart from correcting
for the instrumental response, the overall intensity is as measured and has
not been shifted.}%
\label{Fig:1}%
\end{figure}

In most systems, such as the cuprates and pnictides, incident light energies
are typically high enough to directly excite interband charge transfer and
thus the effective mass approximation may be not valid for these systems. When
this happens, strong resonance effects occur and the intensity of the
electronic Raman scattering varies greatly with the incident photon energy, as
for instance in cuprates.\cite{Phonons} However, as shown in Fig. \ref{Fig:1}
in the pnictides the resonance effects are weak and thus off-resonance
conditions for the scattering amplitude applies. Therefore we assume that the
effective mass approach holds for the pnictides and examine the full band and
momentum dependence of the scattering amplitudes as a way of pinpointing the
gap extrema structure.

In the superconducting state, the Raman response for $B_{1g}$ and $B_{2g}$
symmetries is given as a weighted average of the Tsuneto-Maki function
$\lambda_{n}$ over the FS for each band $n$\cite{Multiband},
$
\chi_{\gamma,\gamma}(\Omega)=\sum_{n,\mathbf{k}}\gamma_{n}^{2}(\mathbf{k}%
)\lambda_{n}(\mathbf{k},\Omega)
$,
with the Raman vertices $\gamma$ given in Eq. \ref{basic}. The Tsuneto
function is
\begin{equation}
\lambda_{n}(\mathbf{k},\Omega)=\tanh\frac{E_{n}(\mathbf{k})}{2k_{B}T}%
\frac{4\mid\Delta_{n}(\mathbf{k})\mid^{2}/E_{n}(\mathbf{k})}{4E_{n}%
^{2}(\mathbf{k})-(\hbar\Omega+i\delta)^{2}},\label{Eq:Tsuneto}%
\end{equation}
where $\Delta_{n}(\mathbf{k})$ is the superconducting gap in band $n,$  having
quasiparticle energy dispersion $\epsilon_{n}(\mathbf{k})$, and $E_{n}%
^{2}(\mathbf{k})=\epsilon_{n}^{2}(\mathbf{k})+\Delta_{n}^{2}(\mathbf{k})$.
Enforce particle number conservation, one gets
\begin{equation}
\chi(\Omega)=\chi_{\gamma,\gamma}(\Omega)-\chi_{\gamma,1}(\Omega
)\chi_{1,\gamma}(\Omega)/\chi_{1,1}(\Omega),\label{Eq:backflow}%
\end{equation}
with $\chi_{\gamma,1}(\Omega)=\chi_{1,\gamma,}(\Omega)=\sum_{n,\mathbf{k}%
}\gamma_{n}(\mathbf{k})\lambda_{n}(\mathbf{k},\Omega)$ and $\chi_{1,1}%
(\Omega)=\sum_{n,\mathbf{k}}\lambda_{n}(\mathbf{k},\Omega)$. Therefore, while
for both $B_{1g}$ and $B_{2g}$ each contribution to the Raman vertex can be
separated into contributions from single bands, the $A_{1g}$ contributions mix
mass fluctuations in each band and between bands.

In Ref. \onlinecite{Muschler:2009}, the interplay of the energy gap and the
symmetry dependence of the Raman vertices was considered, where it was argued
that $B_{2g}$ polarizations largely sampled the energy gap on the electron
bands, while $A_{1g}$ was more sensitive to the hole bands in the pnictides.
Here we explore the form for the Raman vertices in more detail using LDA.

The Raman vertices shown in Eq. (\ref{basic}) are directly computable from the
one-electron band structure. We have performed full-potential LAPW
calculations using the WIEN2k package and an experimental crystal structure,
including the spin-orbit interaction. The doping is accounted for in the virtual crystal approximation.
The derivatives of the eigenstates were
calculated numerically using a 31x31x31 $k$-point mesh in the bct BZ. We
neglect any momentum interplay between the Raman vertices and the
superconducting gaps in the Tsuneto-Maki function, and examine just the
character of the Raman vertices on each band for each polarization
geometry.\cite{Krantz} The calculated quantities are derived from evaluating
Eq. \ref{Eq:backflow} as
\begin{equation}
\chi_{\gamma,\gamma}^{sc}=\sum_{n,\mathbf{k}}\gamma_{n}^2(\mathbf{k}%
)\delta\lbrack\epsilon_{n}(\mathbf{k})]-\frac{\left(\sum_{n,\mathbf{k}}\gamma
_{n}(\mathbf{k})\delta\lbrack\epsilon_{n}(\mathbf{k})]\right)^2}{\sum_{n,\mathbf{k}%
}\delta\lbrack\epsilon_{n}(\mathbf{k})]}.
\end{equation}
This can be simplified by writing $\gamma_{n}(\mathbf{k})=\Delta\gamma
_{n}(\mathbf{k})+\bar{\gamma}_{n}$, with $\bar{\gamma}_{n}=\sum_{\mathbf{k}%
}\gamma_{n}(\mathbf{k})\delta\lbrack\epsilon_{n}(\mathbf{k})]/\sum
_{\mathbf{k}}\delta\lbrack\epsilon_{n}(\mathbf{k})]$ the average of the Raman
vertex for band $n$ over Fermi surface sheet $n$. The response can be
separated into an intraband piece, which is the only contribution for $B_{1g}$
and $B_{2g}$,
\begin{equation}
\chi_{intraband}=\sum_{n,\mathbf{k}}[\Delta\gamma_{n}(\mathbf{k})]^{2}%
\delta\lbrack\epsilon_{n}(\mathbf{k})],
\end{equation}
and an interband contribution only appearing for $A_{1g}$ channels $\sum
_{n}N_{n}\bar{\gamma}_{n}^{2}-(\sum_{n}N_{n}\bar{\gamma}_{n})^{2}/\sum
_{n}N_{n}$. The last part can be written as a band-dependent sum:
\begin{equation}
\chi_{interband}=\sum_{n}\frac{N_{n}}{N_{t}}\bar{\gamma}_{n}\sum_{m}N_{m}%
(\bar{\gamma}_{n}-\bar{\gamma}_{m})=\sum_{n}\chi_{n,interband},
\end{equation}
with $N_{t}=\sum_{n}N_{n}$ the total DOS. The results are shown in Table
\ref{table} and are defined as follows:
\begin{align}
N_{n} &  =\sum_{\mathbf{k}}\delta\lbrack\epsilon_{n}(\mathbf{k})] \nonumber\\
\gamma_{n,\alpha\beta} &  =\sum_{\mathbf{k}}[{\partial^{2}\epsilon
_{n}(\mathbf{k})}/{\partial k_{\alpha}\partial k_{\beta}}]\delta\lbrack
\epsilon_{n}(\mathbf{k})]. \nonumber\\
\gamma_{n,\alpha\beta\gamma\delta} &  =\sum_{\mathbf{k}}[{\partial
^{2}\epsilon_{n}(\mathbf{k})}/{\partial k_{\alpha}\partial k_{\beta}}]%
[{\partial^{2}\epsilon_{n}(\mathbf{k})}/{\partial k_{\gamma}\partial
k_{\delta}}]\delta\lbrack\epsilon_{n}(\mathbf{k})],  \nonumber
\end{align}
such that the $B_{1g}$ polarizations select $(\gamma_{n,xxxx}-\gamma
_{n,xxyy})/2$, $B_{2g}$ selects $\gamma_{n,xyxy}$. The unscreened intraband
part for $A_{1g}$ is given by $(\gamma_{n,xxxx}+\gamma_{n,xxyy})/2$, which is
reduced with screening to $(\gamma_{n,xxxx}+\gamma_{n,xxyy})/2-N_{n}%
\bar{\gamma}^{2}$.

\begin{table}[h]
\begin{center}%
\begin{tabular}
[c]{|l||c|c|c|c|r|}\hline
& h1 & h2 & h3 & e1 & e2\\\hline\hline
$N_{n}$ & 10.85 & 16.4 & 12.97 & 12.33 & 10.84\\\hline
$(\gamma_{xx}+\gamma_{yy})/2 $ & -18.2 & -9.9 & -12.3 & 9.1 & 19.7\\\hline
$\bar\gamma$ & -1.7 & -0.6 & -1.0 & 0.74 & 1.8\\\hline
$(\gamma_{xxxx}+\gamma_{xxyy})/2$ & 39.2 & 9.4 & 15.9 & 19.7 & 47.8\\\hline
$A_{1g}$ (intraband) & 7.8 & 2.6 & 2.9 & 15.1 & 12.7\\\hline
$A_{1g}$ (interband) & 28.6 & 3.6 & 7.7 & 8.6 & 38.9\\\hline
$B_{1g}$ & 2.2 & 4.3 & 6.3 & 32.8 & 8.5\\\hline
$B_{2g}$ (averaged) & 4.8 & 8.5 & 5.2 & 21.8 & 13.8\\\hline
\end{tabular}
\end{center}
\par
\vspace{-0.1cm}\caption{Values of the Raman vertices and related quantities
including spin-orbit coupling around the hole ($h1,h2,h3$) and electron
($e1,e2$) Fermi surfaces, on a per cell 
 basis, in Ry units (energy in Ry, distance in Bohr
radii, momentum in inverse Bohr radii). Here $h1...h3/e1...e2$ denote the
outermost...innermost hole/electron barrel Fermi surface, centered at $\Gamma
$/$M$ in the Brillouin zone, as indicated in Fig. \ref{Fig:3}. }%
\label{table}%
\end{table}


The LDA results shown in Table \ref{table} indicate that both $B_{1g}$ and
$B_{2g}$ polarizations give vertices which are dominated by contributions from
the electron FS, in agreement with the assessment of Ref.
\onlinecite{Muschler:2009}. Moreover we note that $B_{1g}$ vertices are
specifically large only on a single electron FS sheet, given by the outer FS
pocket centered at $M$. For the $A_{1g}$ contribution, while the unscreened
Raman vertex $\gamma_{xxxx}+\gamma_{xxyy}$ has large contributions from the
hole band $h1$ and electron band $e2$, backflow largely cancels the intraband
contribution, leaving a substantial interband contribution deriving largely
from these same bands. These results support the conjecture based on symmetry
in Ref. \onlinecite{Muschler:2009} that the hole bands would be probed for
$A_{1g}$ configurations more so than for $B_{1g}$ and $B_{2g}$.

The LDA results indicate that the $A_{1g}$ interband portion
involves also a strong contribution from scattering between the electron
band $e2$ and the hole band $h1$, coming from different signs of the Raman
vertices in the two bands. As discussed in Ref. \onlinecite{Multiband} the
divergence at twice the gap edge which may appear in crossed polarization
orientations (such as $B_{1g,2g}$) is removed in single band superconductors
for $A_{1g}$ symmetries due to the gauge invariant backflow which screens
uniform charge displacements over many unit cells. In multiband systems, such
as the pnictides, the superconducting response for $A_{1g}$
symmetries may be written as\cite{Multiband}
\begin{equation}
\chi_{A_{1g}}^{scr}(\Omega)=\sum_{n,\mathbf{k}}\left(  \gamma_{n}%
(\mathbf{k})-\bar{\gamma}(\Omega)\right)  ^{2}\lambda_{n}(\mathbf{k}%
,\Omega),\label{Eq:chiscr}%
\end{equation}
with $\lambda_{n}$ given by Eq. (\ref{Eq:Tsuneto}). The Tsuneto-Maki function
$\lambda_{n}$ possesses divergences when $\Omega=2\left\vert \Delta
_{n}(\mathbf{k})\right\vert $. $\bar{\gamma}(\Omega)$ is another way of
representing the frequency-dependent backflow in the superconducting state for
$A_{1g}$ symmetries:
\begin{equation}
\bar{\gamma}(\Omega)=\sum_{n,\mathbf{k}}\gamma_{n}(\mathbf{k})\lambda
_{n}(\mathbf{k},\Omega)/\sum_{n,\mathbf{k}}\lambda_{n}(\mathbf{k},\Omega).
\end{equation}
Each separate divergence occuring at the gap edge in each band in $\lambda
_{n}(\mathbf{k},\Omega)$ weighs exactly those momentum points in $\gamma
_{n}(\mathbf{k})$ for which the divergence occurs, and via Eq. \ref{Eq:chiscr}
the divergence is cancelled. This holds for each divergence coming from each
band $n$, and therefore we conclude that no divergence may appear in $A_{1g}$
symmetries for multiband systems. This conclusion holds only for Cooper pair
creation via light scattering but does not consider additional divergences
coming from charge or spin collective modes.\cite{CKD,Klein}

This predicted absence of an $A_{1g}$ divergence is in general agreement in
the observed data on the pnictides, where no clear $2\Delta$ scattering peak
is seen for $A_{1g}$ scattering geometries. While sharp collective modes may
appear due to the oscillation of pairing amplitudes in the different bands for
$A_{1g}$ symmetries\cite{Chubukov}, no clear indication of such a mode has
been found. Similarly, although collective modes have been predicted to also
occur in $B_{2g}$ symmetries due to final state excitons having orthogonal
symmetry to the superconducting pair state\cite{Scalapino}, it is still an
open question whether these modes can be uncovered in the data.

\begin{figure}[t]
\includegraphics[width=.42\columnwidth]{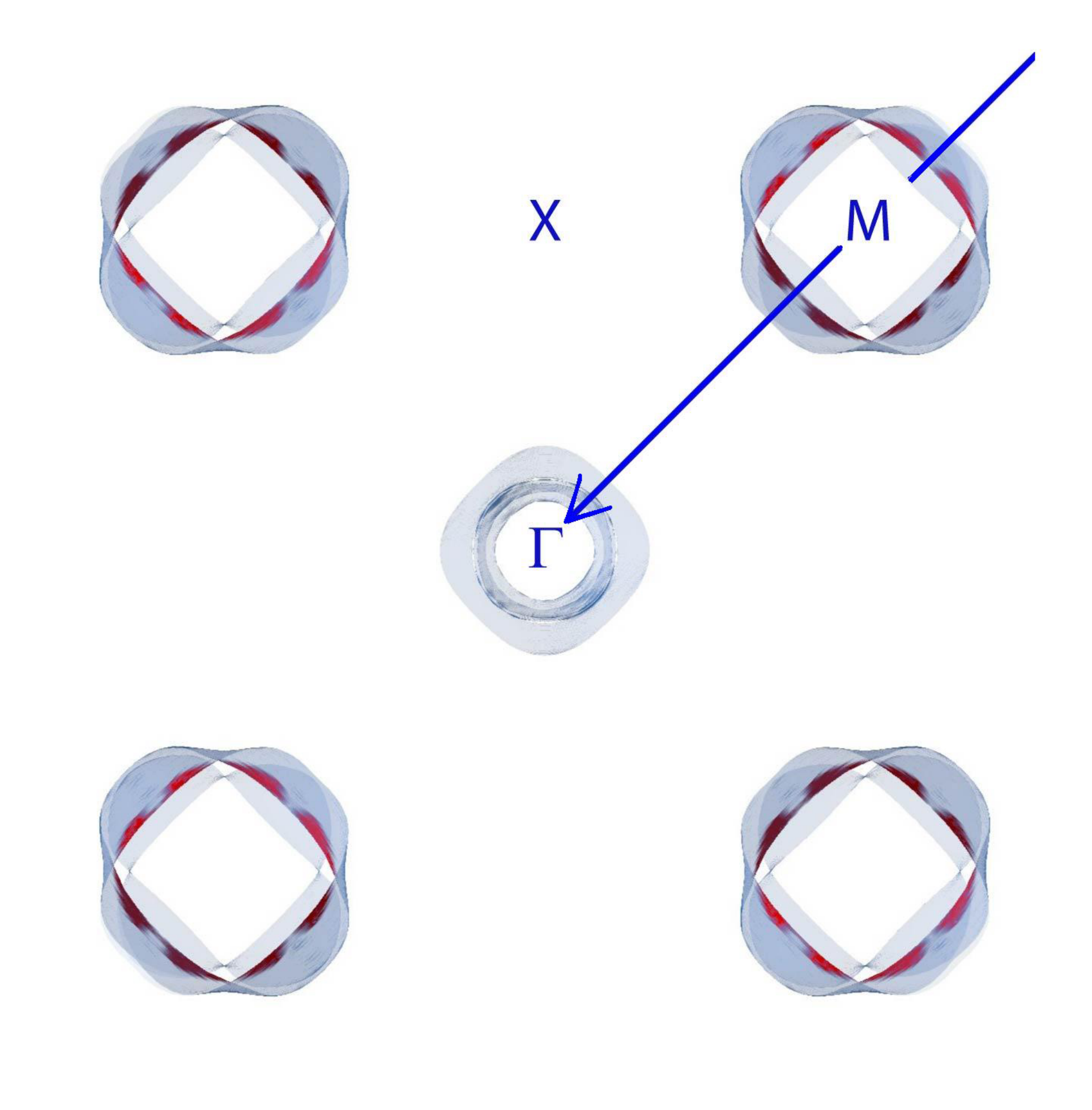}
\includegraphics[width=.56\columnwidth]{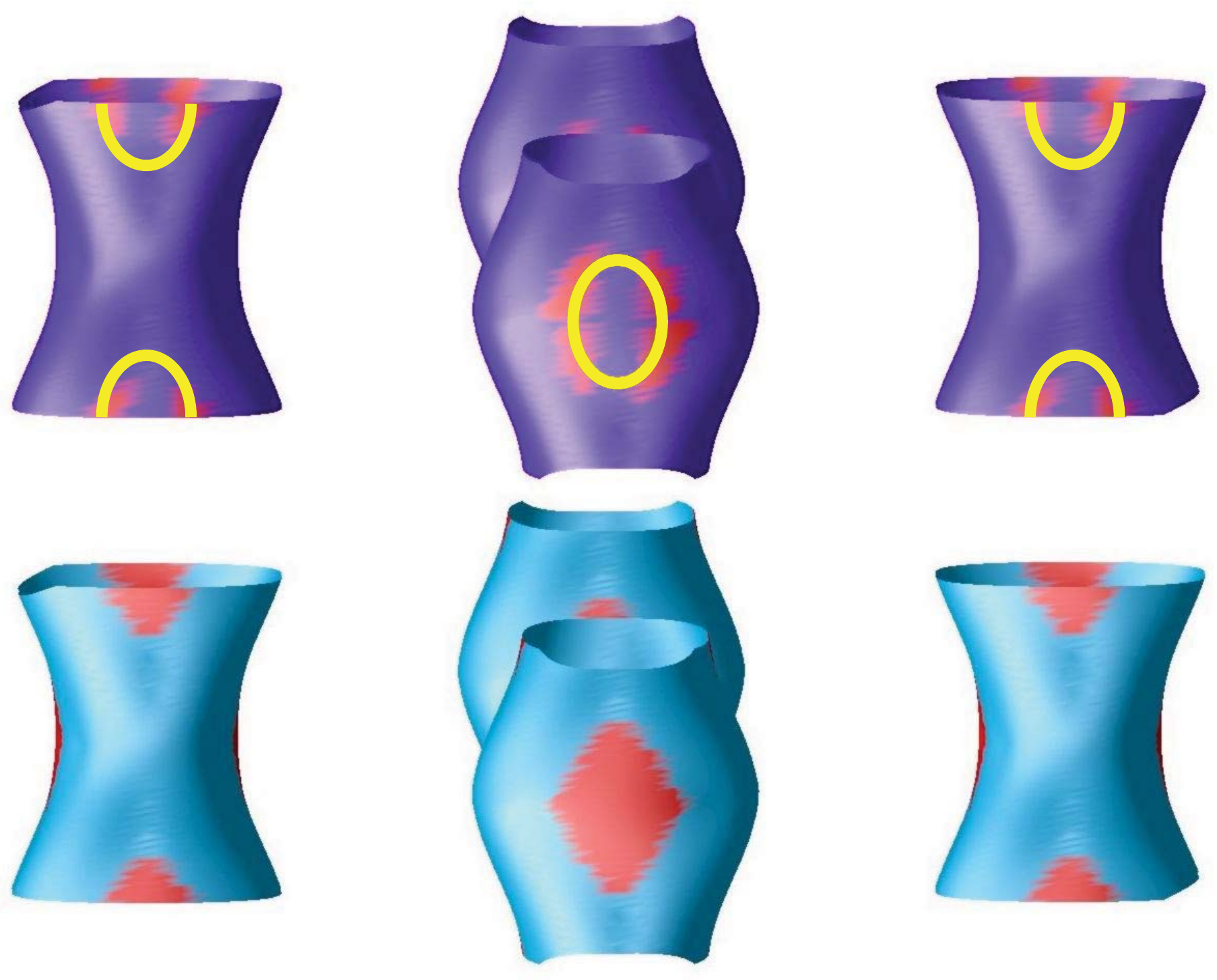}
\caption{(Color online) Left:
Raman vertices for $B_{1g}$ symmetry,
evaluated around each Fermi surface sheet
projected onto the $x-y$ plane, that is, looking down the $k_{z}$-direction. 
The parts
of the Fermi surface where the value of $B_{1g}$ Raman vertex is larger than
the average are shaded red. The gold ellipses mark the area where the deep gap minima or nodes
are expected.
Right top: The same, viewed along the
$M-\Gamma$ line, as shown by the arrow in the left panel;
 only the outer electron FS pockets are shown.
Right bottom: The same view, but now the parts where the Fermi velocity is 
larger than 60\% of its maximal
value are red.}
\label{Fig:3}%
\end{figure}

The fact that in the experiment hardly any effect of superconductivity is
observed in the $B_{1g}$ polarization, while the calculations predict a
noticeable coupling with the outer electronic FS, suggests that this coupling
is not uniform, but is small wherever the gap is sizeable, and where the
coupling is large the gap is small. In order to test this hypothesis, we have
plotted the effective mass asymmetry, appearing in $B_{1g}$ polarizations as
$|\partial^{2}\epsilon_{n}(\mathbf{k})/\partial k_{x}^{2}-\partial^{2}%
\epsilon_{n}(\mathbf{k})/\partial k_{y}^{2}|,$ in Fig. \ref{Fig:3}.
 We observe that, first, this difference is small on all FSs
except the outer e-barrel, and, second, it is also small on the most of that
FS as well, except on a small spot located at an angle varying from 10 to
15$^{o},$ depending on $k_{z}$. As Fig. \ref{Fig:3} shows,
this is consistent with gap nodes or deep gap minima, forming elliptical loops
on the outer electron FS centered around the ($\xi,\xi,0)$ points, where the
$\Gamma-M$ line intersects the electron pockets (indicated by the ellipse 
in Fig. \ref{Fig:3}). This can be compared with model calculations by Graser
$et$ $al.$\cite{Kemper} who found that the pairing interaction is strong in
the $Z$ plane ($k_{z}=\pi/c)$ plane and weaker in the $\Gamma$ plane. In their
calculations this led to a nodal loops similar to ours, but located at the
hole FS. One should be, however, careful, when making a quantitative, as
opposed to qualitative, comparison, with these calculations, because, as
discussed above, Graser $et$ $al.$ worked in an unfolded BZ wher the
folding-induced 3D dispersion was emulated through enhance $k_{z}$ hopping
parameters. 

It is also consistent with the thermal conductivity data\cite{Matsuda}
(collected on another 122 material), which indicate that there are gap nodes
present in the system, yet the specific heat does not seem to be affected by
them. These data have been interpreted in terms of ultraheavy hole bands that
contribute at least 80\% to the DOS, and ultralight electron bands that
dominate the transport. However, the former assumption seems a bit too strong
to be realistic. Yet if in BaFe$_{2}$(As,P)$_{2}$ the gap structure resembles
the one that we propose here for BaFe$_{1.88}$Co$_{0.12}$As$_{2},$ so that the
gap nodes exist only for a finite range of $k_{z},$ the rest of the electronic
bands, fully gapped, add their contribution to the hole band and together
provide 80\% or more of the total specific heat. In fact, incidentally, the
Fermi velocity is the largest close to the \textquotedblleft hot
spots\textquotedblright\ that we have identified as locations of gap nodes, in
agreement with the idea that the transport is dominated by near-nodal parts of
the FS (Fig. \ref{Fig:3}).

One can ask a question, why most of the Fermi surface, except for relatively
small \textquotedblleft hot spots\textquotedblright,\ shows such weak coupling
in the $B_{1g}$ channel. This is actually easy to address. Let us start with a
simple model, a 2D parabolic electronic band. By symmetry, the principal axes
of the corresponding elliptical FS are along (11) and (1$\bar{1}).$ For the
moment, we will use the center of this FS as the origin and the principal axes
as coordinates. Then $\epsilon_{1,2}=\mu_{1,2}\bar{k}_{x}^{2}+\mu_{2,1}\bar
{k}_{y}^{2}.$ The Raman vertex in question appears like a $B_{2g}$ vertex in
this basis and is proportional to $\partial^{2}\epsilon_{n}(\bar{k}_{x}%
,\bar{k}_{y})/\partial\bar{k}_{x}\partial\bar{k}_{y}=0.$ Thus in this model
this band does not couple in the $B_{1g}$ channel at all.

However, this is not necessarily true if one takes into account the
downfolding of the electron pockets in the crystallographic unit cell. After
the downfolding the two FSs simply intersect. In our model, as well as in the
real 11 and 1111 crystallography, they cross along the $\bar{k}_{x}=0$ and
$\bar{k}_{y}=0$ directions, and, because they have different parities, there
is no hybridization between them and the newly formed degeneracy is not lifted
(although it would be lifted by the spin-orbit coupling). On the other hand,
in the 122 symmetry the intersections occur at general points in the Brillouin
zone so that no symmetry prevents them from hybridization, thus forming an
inner barrel and an outer barrel ($e1$ and $e2$) even without the spin-orbit.
 To gain more insight into
this problem we introduce hybridization into our model, and write
the Hamiltonian near the anticrossing point as $H_{11}=\epsilon_{1}%
(\mathbf{\bar{k}),}H_{22}=\epsilon_{2}(\mathbf{\bar{k}),}H_{12}=H_{21}=V$,
where $V$ is a hybridization matrix element. The inverse masses of the
resulting two bands are
\begin{equation}
\left\vert \frac{\partial^{2}\epsilon_{\pm}}{\partial\bar{k}_{x}\partial
\bar{k}_{y}}\right\vert =\frac{8V^{2}|(\mu_{1}-\mu_{2})^{2}\bar{k}_{x}\bar
{k}_{y}|}{[(\mu_{1}-\mu_{2})^{2}(\bar{k}_{x}^{2}-\bar{k}_{y}^{2})^{2}%
+4V^{2}]^{3/2}}.
\end{equation}
This is essentially an approximate expression of a delta function centered at
$\bar{k}_{x}\pm\bar{k}_{y}=0$ with the width $V$ and strength $|\mu_{1}%
-\mu_{2}|.$ Thus in this simple model we have vanishing coupling everywhere
except hot spots where the bands would cross if the As sublattice would not
have a lower symmetry than the Fe one.

This simplified picture only qualitatively applies to Ba122. The bands there
are not exactly parabolic, and, most importantly they cross not at the
$\bar{k}_{x},\bar{k}_{y}=0$ lines, but at general points that can be
considerably removed from these line, depending on $k_{z}$. Yet our model
demonstrates qualitatively how the \textquotedblleft hot
spots\textquotedblright\ for this polarization appear on the FS, and why they
are so small in size.

In summary, in this paper we have paired experimental data on the electronic
Raman scattering in the superconducting state of Ba(Fe$_{0.94}$Co$_{0.06})_{2}$As$_2$
with the first principle calculations of the effective mass fluctuations. We
found that a 2D TB model in the unfolded Brillouin zone is not
sufficient to capture all the mass fluctuation. The reason is that because of
the body-centered symmetry of the corresponding crystal structure downfolding
the electron bands results in hybridizations and substantial band
anticrossings, which in turns produces large mass fluctuation in a small part
of the Fermi surface. Our results suggest that closed nodal loops (rather than
vertical or horizontal nodal lines) exist in this compound near the
intersection of the outer electron Fermi surface with the $\Gamma-$M line. It is likely that the
size and the very existence of these loops is doping-dependent, but this
is beyond the scope of this paper.

{\it Acknowledgements:} I.I.M. acknowledges support from the Office of Naval Research. R.H acknowledges 
support by the DFG under grant numbers HA 2071/3 and HA 2071/7 via Research Unit FOR538 and priority Program SPP1458, respectively. R.H. and T.P.D. acknowledge support from BaCaTeC. The work at SLAC and Stanford University is supported by the Department of Energy, Office of Basic Energy Sciences under 
contract DE-AC02-76SF00515.

\end{document}